\documentstyle[12pt,epsfig]{article}
\newcommand{\bm}{\bibitem}

\newcommand{\ra}{\rightarrow}
\def\be {\begin{equation}}
\def\ee {\end{equation}}
\def\bea {\begin{eqnarray}}
\def\eea {\end{eqnarray}}
\def\nn {\nonumber}
\def\Ekst {E_k^{\ast}}
\def\cI {{\cal{I}}}
\def\om {\omega}

\begin{document}

\begin{center}
\Large{ Matter induced charge symmetry breaking and pion form factor in 
nuclear medium }
\vskip 0.2in
{Pradip Roy$^a$, 
Abhee K. Dutt-Mazumder$^a$, Sourav Sarkar $^b$, Jan-e  Alam$^b$}

\vskip 0.2in
a) Saha Institute of Nuclear Physics, 1/AF Bidhannagar, Kolkata, India\\
b) Variable Energy Cyclotron Centre, 1/AF Bidhannagar, Kolkata, India
\end{center}

\vskip 0.2in
\begin{abstract}
Medium modification of pion form factor has been evaluated in asymmetric
nuclear matter (ANM). It is shown that both  the 
shape and the pole position of the pion form factor in dense asymmetric
nuclear matter is different from its vacuum counterpart with
$\rho$-$\omega$ mixing. This is due to the density and asymmetry 
dependent $\rho$-$\omega$ mixing which could even dominate over its
vacuum counterpart in matter. Results are presented for arbitrarily
mixing angle. Effect of the in-medium pion factor
on experimental observables {\it e.g.}, 
invariant mass distribution of lepton pairs has been demonstrated.
\end{abstract}

{25.75.Dw; 21.65.+f; 14.40.-n}


\section{Introduction}
The pion electromagnetic form factor, $F_\pi(Q^2)$
in vacuum shows that the physical $\rho$ meson is
not a pure isospin eigenstate and it can mix with the $\omega$ meson
\cite{connell97}.
In vacuum this mixing amplitude can be determined by measuring
$F_{\pi}(Q^2)$ which, although dominated by the $\rho^0$ pole,
shows a kink near the $\omega$ meson mass. Such mixing (after electromagnetic
correction) implies that the charge symmetry is broken at the most fundamental 
level in strong interaction through the small mass difference between
up and down quarks in the QCD Lagrangian. Consequently, 
the physical $\rho$ and $\omega$ mesons that we deal with are admixtures of
the corresponding isospin eigenstates. At the hadronic level
this mixing can be understood in terms of neutron-proton mass difference
in effective models \cite{piekarewicz93}.

$\rho$-$\omega$ mixing has important and interesting consequences. 
It plays a crucial role in generating contributions to few body charge
symmetry violating observables \cite{miller06}. The
$\rho$-$\omega$ mixing amplitude is 
determined from $e^+e^-\ra \pi^+\pi^-$ by measuring the pion form
factor in the interference region \cite{connell97}.
With the extracted value of the mixing amplitude one is able to explain
a number of observables namely, the non-Coulombic binding energy
difference (Nolen Schiffer anomaly~\cite{okamoto64,nolen69}) 
of $A = 3$ (mirror) nuclei, significant
contributions to the $np$ asymmetry at 183 MeV and non-negligible
contributions to the difference of $nn$ and $pp$ scattering lengths
\cite{miller90}.

The mixing of different isospin states will be modified in matter. 
Such medium effects have recently been 
investigated by several authors~\cite{abhee96,broni98,subhra06}.
Unlike neutron-proton
mass difference, which is responsible for $\rho$-$\omega$ mixing
in free space, the mixing in matter can be induced if the 
neutron-proton densities are different.
This happens even if the Hamiltonian preserves the isospin
symmetry {\it i. e.}, if $M_n = M_p$, akin to the `spontaneous symmetry 
breaking' driven by the $n\leftrightarrow p$ asymmetric ground state.
It has been shown in Ref.\cite{abhee96}  
that the density dependent mixing is of similar magnitude as the usual vacuum 
mixing at normal nuclear matter density. Subsequently, Broniowski and
Florkowski, showed that the mixing could be significantly large in
$Pb$ like nuclei~\cite{broni98}. These studies were limited to hadronic
models. In contrast, such density dependent $\rho$-$\omega$ mixing 
was studied within the framework of QCD sum-rule in
Ref.~\cite{abhee01} with similar results and conclusions. 
The sum rule calculation has recently been improved and extended
to study the density dependent
$\rho$-$\omega$ amplitude and its observable implications in
heavy ion collisions~\cite{zschocke04}. 

In the present paper we revisit the problem
within the framework of Walecka model \cite{serot86}.
We discuss the possible consequences 
in presence of a scalar mean field, an effective way to incorporate 
nucleon-nucleon interactions. Moreover, as in dense matter with large
neutron-proton density asymmetry, the mixing angle could be quite high,
we present results valid for arbitrary mixing angles.
As an application, 
the possible modification of pion form factor in 
ANM and the $\pi^+\pi^- \rightarrow e^+e^-$ annihilation 
cross section at various densities and asymmetries have also been discussed.
This is relevant for
the dilepton production in relativistic heavy ion collisions. Therefore, the
present investigation has direct relevance for the study of compressed
baryonic matter (CBM) expected to be produced in heavy ion collisions
at GSI energies~\cite{CBM}.

Apart from $\rho-\omega$ mixing in asymmetric matter $\rho$ can
also mix with the $\sigma$ in the same way as $\omega-\sigma$ 
mixing~\cite{wolf}.
We have discussed such a possibility in Appendix I where it is argued
that for matter with small neutron-proton asymmetry such possibilities
could be ignored.

The paper is organized as follows. In the next section the formalism 
is set forth. In section 
3 the relations between dilepton production cross section and form factor
with mixing has been presented. Results are discussed in section 4 
while section 5 is devoted to summary. Mathematical expressions 
relevant for the calculation of 
pion form factor with matter induced mixing and
$\rho$ and $\omega$ meson self energies in dense nuclear matter are 
relegated to the appendix. 

\section{Formalism}
\begin{figure}[htb]
\begin{center}
\epsfig{figure=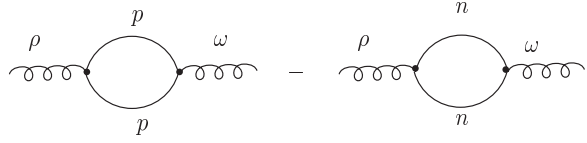, height=2.0 cm} 
\caption{$\rho$ and $\omega$ nucleon loop }
\label{fig1}
\end{center}
\end{figure}

The Lagrangian describing $\rho$ - $\omega$ meson-nucleon 
interaction is given by,

\bea
{\cal L}&=&\bar{\psi}(\tau_a\Gamma_{a,\mu} \rho^\mu_a)\psi
+\bar{\psi}(\Gamma_{\omega,\mu}\omega^\mu)\psi,\nn\\
\Gamma_{v,\mu}&=&g_v \left [ 
\gamma_\mu-\frac{\kappa_v}{2{\bar{M}}}\sigma_{\mu\nu}
\partial^\nu \right ].
\eea
Here ${\bar{M}} = (M_p+M_n)/2$ where the subscript $p$ ($n$)
stands for proton (neutron). Clearly, due to the Pauli
matrix $\tau_3$, $\rho$ meson couples to neutron with a 
negative sign in contrast to the $\omega$ meson while with proton both
couple with the same sign.
This gives rise to mixing that vanishes in the limit,  
$M_n = M_p$ which is illustrated in 
Fig.\ref{fig1}. 
\vskip 0.5cm
Here the mixing is generated by $N\bar{N}$ loop. The vacuum
$\rho$-$\omega$ mixing in this approach was first calculated in 
Ref.\cite{piekarewicz93}.
We calculate similar mixing in ANM which, besides $n$-$p$ mass difference,
would also depend both on baryon density $(\rho_B)$ and the 
asymmetry parameter $\alpha=(\rho_n-\rho_p)/(\rho_n+\rho_p)$.

To determine the density and asymmetry dependent mixing amplitude we
evaluate these loops at finite density which
are characterized by the following density dependent polarization functions :
\begin{equation}
\Pi_{\mu\nu}^{\alpha\beta}=\frac{-i}{(2\pi)^4}
\int {d^4}K~{\rm Tr}[i\Gamma^\alpha_\mu iG(K+Q)
{i\bar\Gamma^\beta_\nu} iG(K)]
\end{equation}
where,
\begin{equation}
G(K) = G_F(K)  + G_D(K)
\end{equation}
\begin{equation}
G_F(K) =
(K_\mu \gamma^\mu +M^\ast)[\frac{1}{K^2-M^{\ast2}+i\epsilon}]
\end{equation}
\begin{equation}
G_D(K) = (K_\mu \gamma^\mu + M^\ast) [\frac{i\pi}{E_k^{\ast}}
\nonumber\\ \delta (k_0-E_k^\ast )\theta(k_F-|\vec k |)]
\end{equation}
Here $G_F$ denotes free nucleon propagator while $G_D$ represents
the density dependent part which forbids on mass shell nucleon propagation 
in matter due to
Pauli blocking \cite{serot86}. As the propagator now explicitly involves the 
Fermi
momentum, contributions will be different for the neutron and proton
loops if their densities (and hence Fermi momenta) 
are different. It is to be noted that the
polarization function or the mixing amplitude will have two parts,
one is like the polarization tensor with the free nucleon mass
replaced by its effective mass called the free part representing
nucleon-antinucleon excitations (Dirac sea) and the other one is density
dependent part relevant for the scattering from the Fermi sphere. 
We discuss them separately.

\subsection{Density dependent part}

The density dependent piece of the mixed polarization 
($\rho$-$\omega$) due to $p$-$p$ or $n$-$n$ excitations 
is generically given by
\begin{eqnarray}
\Pi^D_{\mu\nu}(k_F,M^*,q_0,|q|)
&=& \frac{g_\rho g_\omega \pi}{(2\pi)^4}\int\frac{{d^4}K}{E_k^{\ast}}
\delta (k^0-E_k^{\ast})
\nonumber\\
&&\theta(k_F-\mid \vec k\mid)
\times \left[\frac{{\it {T}}_{\mu\nu}(K-Q,K)}{(K-Q)^2-M^{\ast 2}}
\right.\nonumber\\
&&\left.+\frac{{\it {T}}_{\mu\nu}(K,K+Q)}{(K+Q)^2-M^{\ast 2}}\right]
\end{eqnarray}
where $T_{\mu\nu}$ represents relevant traces as in Ref.\cite{chin77}.

The $\Pi_{\mu\nu}^D(q)= \Pi_{\mu\nu}^{vv}+\Pi_{\mu\nu}^{vt}$ functions in 
this case are as follows
\begin{equation}
\Pi_{\mu\nu}^{vv}
(k_F,M^*,q_0,|q|)
=\frac{g_\rho g_\omega}{\pi ^3}\int_0^{k_F}\frac{{d^3k}}{E_k^{\ast}}
\frac{Q^2{\cal K}_{\mu\nu}-Q_{\mu\nu}(K\cdot Q)^2}{Q^4-4(K\cdot Q)^2}
\end{equation}
                                                                                
\bea
\Pi_{\mu\nu}^{vt} (k_F,M^*,q_0,|q|)
&=&-\frac{g_\rho g_\omega}{\pi ^3}
(\frac{(k_\rho)M^\ast}{8{\bar M}})2Q^4Q_{\mu\nu}\nonumber\\
&&\times\int_0^{k_F}\frac{{d^3k}}{E_k^{\ast}}
\frac{{1}}{Q^4-4(K\cdot Q)^2}
\eea
                                                                                
\vskip 1 true cm
\noindent where ${\cal K}_{\mu\nu}=(K_\mu-\frac{K.Q}{Q^2}Q_\mu)
(K_\nu-\frac{K.Q}{Q^2}Q_\nu)$, $Q_{\mu\nu}=(-g_{\mu\nu} +
\frac{Q_{\mu}Q_{\nu}}{Q^2})$ and $E_k^*=\sqrt{k^2+M^{* 2}}$.  
In the above expressions, for proton 
(neutron) loop we substitute  
$M$ and $k_F$ with $M_p$ ($M_n$) and 
$k_F^p$ ($k_F^n)$ respectively. 
Moreover, at this point
it might be recalled that for a vector meson moving in nuclear matter 
the longitudinal ($L$) transverse ($T$) polarization tensors are different
because of ${\cal{K}}_{\mu\nu}$,
unlike the vacuum part which is  proportional to $Q_{\mu\nu}$. The 
$L$ and $T$ modes are
constructed as $\Pi_L=-\Pi_{00}+\Pi_{33}$ and $\Pi_T=\Pi_{22}=\Pi_{11}$,
where the meson momentum $Q=(q_0,0,0,|q|)$
(see Appendix for details).

The mixing amplitude is characterized by $ \Pi^{\rho\omega}_{L,T}$ which involves
scattering from the neutron and proton Fermi spheres :
\bea
\Pi^{\rho\omega}_{L,T}=
\Pi_{L,T}^{\rho\omega}(k_F^p,M_p^*,q_0,|q|)-
\Pi_{L,T}^{\rho\omega}(k_F^n,M_n^*,q_0,|q|).
\eea
The negative sign arises
because of the $\tau_3$ in the $\rho$-$NN$ interaction.

The pure part of the polarization can be obtained by taking appropriate
vertex factor like $g_\omega$ or $g_\rho$ for both the vertices. Accordingly,
the total is given by a {\it{sum}} over the neutron and proton loops 
instead of the {\it{difference}}. 
\bea
\Pi^{\rho\rho}_{L,T}=
\Pi_{L,T}^{\rho\rho}(k_F^p,M_p^*,q_0,|q|)+
\Pi_{L,T}^{\rho\rho}(k_F^n,M_n^*,q_0,|q|)
\eea
and similarly $\rho\rightarrow \omega$ gives results for the $\omega$ meson
polarization function.
We take $g_\omega=10.1$, $\kappa_\omega=0$ and
$g_\rho=2.63$, $\kappa_\rho=6.0$~\cite{hatsuda96} in numerical computations.

\subsection{Free part}
The vacuum part will also give rise to mixing which is same as Ref.\cite{piekarewicz93}
with $n$ and $p$ mass $M_{n,p}$ replaced by the in-medium masses, $M^*_{n,p}$.
In Walecka model this is determined
from the following self-consistent condition \cite{serot86},
\bea
M_{n,p}^*=M_{n,p}- \frac{g_\sigma^2}{m_\sigma^2} (\rho_p^s + \rho_n^s),
\label{Mstar}
\eea
where $\rho^s_i$ ($i=p,n$) represent scalar densities given by
\bea
\rho^s_i=\frac{M_i^*}{2\pi^2} \left[E^*_i k_F - M_i^{* 2}
\ln \left ( \frac{E^*_i + k_F}{M_i^*}\right )\right ].
\label{emass}
\eea

The free part of the polarization tensor can be written as, 
\bea
\Pi_F^{\mu \nu} = (-g^{\mu \nu}+Q^{\mu} Q^{\nu}/Q^2)\Pi_F(Q^2).
\eea
 The mixing contributions to $\Pi_F$ are given by :
\bea
\Pi_{F,vv}^{\rho \omega}& =& \frac{g_{\rho}g_{\omega}}{2\pi^2}\,Q^2
\int_{0}^{1}\,dz\,z(1-z)\,\left[\ln\frac{M_p^{\ast 2}-Q^2\,z(1-z)}
{{\bar M}^2-Q^2\,z(1-z)}\right.\nonumber\\
&&\left.\frac{}{}- (p\ra n)\right ],
\eea

\bea
\Pi_{F,vt}^{\rho \omega}& =& \frac{g_{\rho}g_{\omega}}{2\pi^2}\,
\frac{\kappa_{\rho}}{4}Q^2
\left[\frac{M_p^{\ast}}{M_p}
\int_{0}^{1}\,dz\ln\left\{\frac{M_p^{\ast 2}-Q^2\,z(1-z)}
{{\bar{M}^2}-Q^2\,z(1-z)}\right\}\right.\nonumber\\
&&\left.\frac{}{}-(p\ra n)\right].
\eea

The pure $\rho$ ($\omega$) meson self-energies for the vector-vector,
vector-tensor and tensor-tensor parts are given by~\cite{hatsuda96}
\bea
\Pi_F^{\rho\rho(\omega\omega)}& =& \frac{g_{\rho(\omega)}^2}{2\pi^2}\,Q^2\sum_{i=p,n}
\left[I_1^{(i)}\frac{}{} \right. \nn\\
&&\left.+\frac{\kappa_{\rho(\omega)}}{2M_i}M_i^{\ast}I_2^{(i)} \right.\nn\\
&&\left.+\frac{1}{2}\left(\frac{\kappa_{\rho(\omega)}}{2M_i}\right)^2
(Q^2 I_1^{(i)}+M_i^{\ast 2} I_2^{(i)})\right],
\eea
where
\be
I_1^{(i)}=\int_{0}^{1}\,dz\,z(1-z)\,\ln\left[\frac{M_i^{\ast 2}-Q^2\,z(1-z)}
{M_i^2-Q^2\,z(1-z)}\right],
\ee
\be
I_2^{(i)}=\int_{0}^{1}\,dz\,\ln\left[\frac{M_i^{\ast 2}-Q^2\,z(1-z)}
{M_i^2-Q^2\,z(1-z)}\right].
\ee

It is to be noted that the free part $\Pi^{\rho\omega}_F$ vanishes in the limit
$M_n=M_p$. We extract the real part of the vacuum mixing amplitude (with
free nucleon mass) to be $\sim -3447 $MeV$^2$ at the omega pole 
and this is consistent with that of Ref.~\cite{gardner98}.
When in-medium nucleon masses are included $\Pi_{\rho \omega}^F(m_{\omega}^2)$
is equal to $-3716$ MeV$^2$ and $-4675$ MeV$^2$ at $\rho_0$ and $2\rho_0$
respectively in symmetric nuclear matter. 


\section{Pion form factor and $\pi^+\pi^- \rightarrow 
{\mathrm{e}}^+{\mathrm {e}}^-$ 
cross section} 

The $\rho$-dominated (unmixed) pion electromagnetic
form factor is given by:
\be
F_\pi(Q^2)=1 - \frac{g_{\rho\pi\pi}}{g_\rho}\frac{Q^2}{Q^2-m_\rho^2+im_\rho\Gamma_\rho}
\ee
in the first form of VMD~\cite{connell97}.

In presence of mixing, the above expression for pion
form factor should be replaced by (see Appendix for detailed derivation)
\begin{eqnarray}
F_{\pi}(Q^2)& =& 1-g_{\rho \pi \pi}\frac{s_{\omega}}{s_{\rho}s_{\omega}-\Pi_{\rho \omega}^2}\frac{Q^2}{g_{\rho}}
-g_{\rho \pi \pi} \tan{\epsilon}\nonumber\\ 
&&\times\frac{s_{\rho}}{s_{\rho}s_{\omega}-\Pi_{\rho \omega}^2}
\frac{Q^2}{g_{\omega}}.
\label{fpi2}
\end{eqnarray}
Here $\tan 2\epsilon=\frac{2\Pi_{\rho\omega}}{s_\rho- s_\omega}$
with $s_{\rho(\omega)}=Q^2-m_{\rho(\omega)}^2-\Pi_{\rho\rho(\omega\omega)}+i
m_{\rho(\omega)}\Gamma_{\rho(\omega)}$.
Here the coupling of the physical $\rho$ and $\omega$ states to the
photon is considered. In this form the $\omega\to \pi^+\pi^-$ decay
is understood to proceed exactly like the $\rho$ but modified by 
the mixing factor $\tan\epsilon$ and $\Pi_{\rho\omega}^2$ as 
appears in the denominator.

In the small mixing limit  
the term quadratic in $\Pi_{\rho\omega}$ in Eq.~\ref{fpi2} can be 
dropped and $\tan\epsilon
\simeq \epsilon$. Thus,
to lowest order in the mixing parameter $\epsilon$, the above expression 
takes the form~\cite{connell97}

\bea
F_\pi(Q^2)=1 - g_{\rho\pi\pi} \frac{1}{s_\rho} \frac{Q^2}{g_\rho}
-g_{\rho\pi\pi}\, \epsilon \frac{1}{s_\omega}
\frac{Q^2}{g_\omega}
\label{fpi1}
\eea


The coupling constants used are $g_{\rho\pi\pi}^2/4\pi \sim 2.9$,
$g_\rho^2/4\pi \sim 2.0 $~\cite{connell97} and 

\bea
\frac{g_\omega}{g_\rho} &=& \sqrt{ 
\frac{m_\omega\Gamma(\rho\rightarrow e^+e^-)}
{m_\rho\Gamma(\omega\rightarrow e^+e^-)}}\nonumber\\
&=&3.5 \pm 0.18.
\eea

In matter, the value of $g_{\rho \pi \pi}$ can acquire a density dependence
due to the coupling with the $\Delta$ excitation as discussed in 
~\cite{broni1,broni2,broni3} in the leading density approximation. It was
shown that the coupling could increase by $\sim 30$ \% near
the $\rho$-pole at nuclear saturation density for finite width of the $\Delta$.
In the zero width approximation it could be as large as double near the 
$\rho$-pole at saturation density. However, as far as our results
are concerned there will be no appreciable change in the dilepton
invariant mass spectrum. The problem of
density dependence of the vertex function is an interesting problem by 
itself and a systematic
approach would be to evaluate the full $\rho$-spectral function in the
presence of mixing. For the present purpose we ignore such effects
in order to bring out the effect of mixed polarization tensor into clearer 
focus.  

The cross section for dilepton production from pion annihilation is 
intimately connected to the density dependent pion form factor which is
given by,
\bea
\sigma(q_0,|q|,\rho_b,\alpha)=\frac{4}{3}\pi \frac{\alpha_{em}^2}{Q^2}
\sqrt{1-\frac{4m_\pi^2}{Q^2}}|F_\pi(q_0,|q|)|^2.
\eea
It is to be noted that unlike in vacuum, the pion form factor in nuclear
medium depends both on $q_0$ and $|q|$.
The dilepton emission rate in terms of the above cross section  is given by
\be
\frac{dR}{dM}=\frac{\sigma(q_0,|q|,\rho_b,\alpha)}{(2\pi)^4} M^4 T K_1(M/T)
\left (1-\frac{4 m_\pi^2}{M^2} \right )
\ee
where $K_1$ is the modified Bessel function of second kind.

It should be mentioned that in deriving the expression for in-medium 
$F_{\pi}(Q^2)$ (Eq.20), the pions have been considered to be on shell.
In relatrivistic heavy ion collisions the main contribution to dilepton
production comes from pion annihilation
where the pions are
generally assumed to be on their mass shells. However, in a medium  
the pion dispersion relation is different from that of vacuum due its
interaction. It would be interesting to extend the above calculation to 
incorporate this feature. This however, goes beyond the scope of the 
present work. 

\section{Results}
\begin{figure}
\centerline{\epsfig{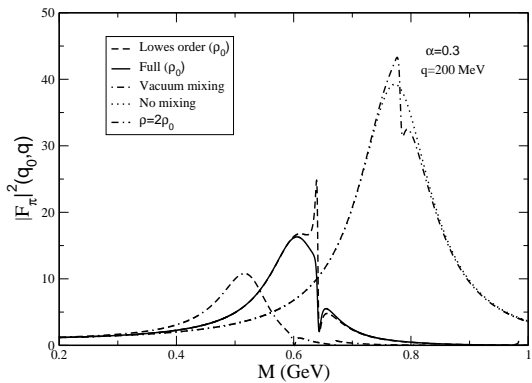}}
\caption{Pion form factor as a function of invariant mass (
$M=\sqrt{Q^2}$) with mean field
including both vacuum and Fermi sea contributions. The dotted and
dot dashed line represent pion form factor in vacuum without and with mixing
respectively. The dashed and solid line depicts the same for $\alpha=0.3$,
and $\rho=\rho_0$ in small angle approximation and for the arbitrary
mixing angle respectively. The dot-dash-dashed line
correspond to the case for arbitrary mixing angle
at $\rho=2 \rho_0$. ``Full (lowest order)'' corresponds to Eq.(20) (Eq.21) }
\label{fig2}
\end{figure}

In Fig.~\ref{fig2}, the in-medium pion form factor  
for $|q|=200$ MeV together with its vacuum counterpart is shown. 
In matter
the pole position shifts towards lower invariant mass indicating the decrease
of $\rho$ and $\omega$ meson masses in matter. It might be mentioned that in
medium, the mass modification is caused by two different mechanisms, {\em viz.}
the scattering from Fermi sphere and excitation of the Dirac vacuum. While
the former gives rise to an increase of their masses, the latter dominates
resulting in an overall reduction. Near the pole,
the mixing amplitude increases by large factors depending upon the value
of the asymmetry parameter $\alpha$. Clearly the density and asymmetry 
parameter dependent mixing is much larger than the mixing due to $n$-$p$ mass 
difference alone. 
In Ref.~\cite{zschocke04} no significant effect of matter induced
mixing on $F_{\pi}$ was observed. However, in the present model
this is found to be substantial which can be attributed to the
tensor interaction. Moreover, we clearly show that the small angle 
approximation are not a good assumption in calculating pion form factor
even for asymmetry relevant for heavy nuclei like $Pb$.

\begin{figure}
\centerline{\epsfig{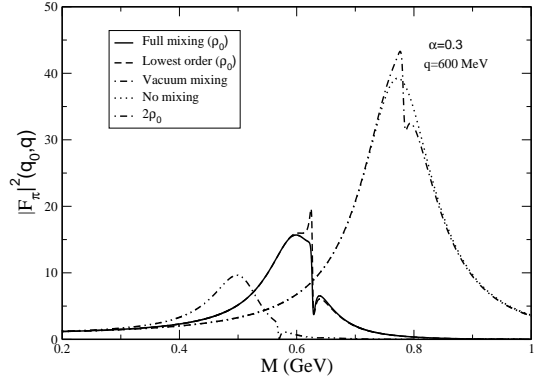}}
\caption{Same as Fig.\protect\ref{fig2} with $q=600$ MeV.} 
\label{fig3}
\end{figure}

Fig.~\ref{fig3} shows results
for $|q|=600$ MeV. Though the qualitative features remain similar,
one can see that the mixing
amplitude depends strongly on the three momentum of the moving meson.
In fact, the mixing is stronger for a slower vector meson. 

\vspace*{.5cm}


\vspace*{4cm}

\vskip 1.5cm 
In Fig.~4, we show the strong asymmetry dependence of the mixing which
shifts the pole of the pion form factor towards lower invariant mass
region. Moreover, we also find that the results are quite sensitive
to the asymmetry parameter ($\alpha$).

As an application of the density and asymmetry
parameter dependent pion form factor, we calculate the pion annihilation
cross section in nuclear matter. 

\begin{figure}[htb]
\centerline{\epsfig{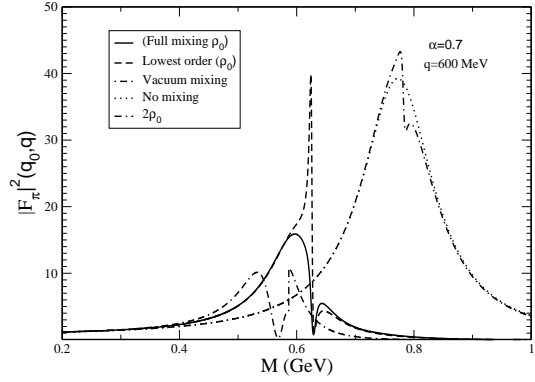}}
\caption{Same as Fig.\protect\ref{fig3} for $\alpha=0.7$. } 
\label{fig4}
\end{figure}
\begin{figure}[htb]
\centerline{\epsfig{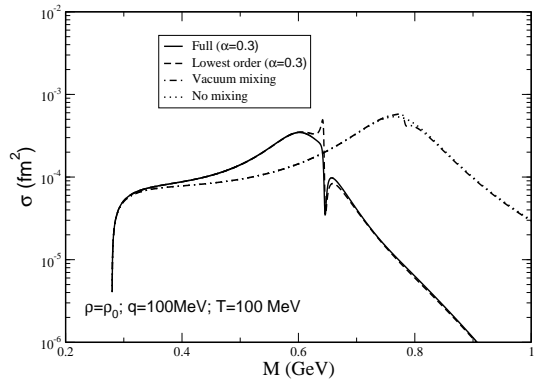}}
\caption{The cross section for $\pi^+\pi^-\rightarrow e^+e^-$. }
 \label{fig5}
\end{figure}
\begin{figure}
\centerline{\epsfig{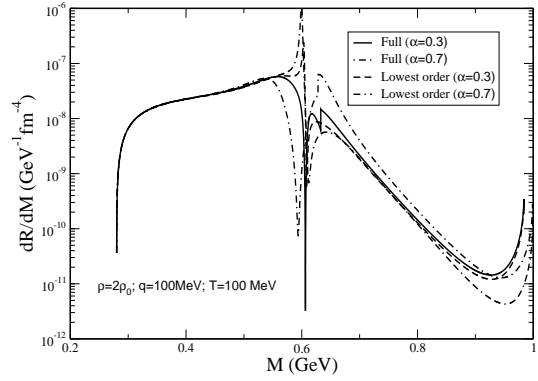}}
\caption{Dilepton production rate with and without mixing at two times
the normal nuclear matter density. }
\label{fig6}
\end{figure}

\vspace*{5cm}
The dilepton
production cross section is directly proportional to the square of the pion
form factor and bears similar qualitative features. 
This is shown in Fig.\ref{fig5}. For completeness, 
we also present results
for the dilepton
production rates for various combinations of density and asymmetry parameter
at $T=100$ MeV in Fig.~\ref{fig6}. Evidently the medium modified pion form 
factor leads to enhanced production of dileptons in the low invariant mass 
region.

\vskip 1.5cm

\vspace*{6cm}

\vspace*{2cm}

\section{Summary}
{
In the present paper we have calculated pion form factor in asymmetric nuclear
matter within the framework of relativistic mean field theory valid for
arbitrarily large mixing angle. It is known
that the vacuum pion form factor shows a shoulder like behaviour near the
$\rho$ meson pole due to its admixture with the isoscalar $\omega$ meson
indicating isospin symmetry violation. In this paper, we have discussed
the possible enhancement of such mixing in asymmetric nuclear matter where
unlike the vacuum case, this symmetry breaking is driven by the ground
state due to the difference of neutron and proton densities. Automatically
this would modify the pion form factor as we have shown. Another interesting
aspect is the shift of pole position. This is related to the presence of
scalar mean field leading to the reduced nucleon mass in matter. In-medium
pion form factor naturally influences the pion annililation cross section
which proceeds through an intermediate $\rho$ meson. Relativistic heavy ion 
collisions at GSI energies offers the unique
opportunity to probe the in-medium pion form factor through dilepton
measurements.  Hence we have also calculated dilepton production rate due
to pion annihilation in matter. In the low mass region the dilepton 
yield is enhanced due to dropping of vector meson mass as well as 
$\rho$-$\omega$ mixing.

We conclude with the comment that density and asymmetry dependent 
$\rho-\omega$ mixing is an interesting problem related to charge
symmetry violation in matter and it
could be worthwhile to extend the formalism developed in the present
work to other approaches such as chiral perturbation theory
and NJL model which have explicit chiral symmetry.
}
\section*{Appendix - I}
\begin{center}
{\Large\bf{Expressions for mixing angles and pion form factor}}
\end{center}
In this section we derive the expression for mixing angle and pion form factor
for arbitrary mixing. In $\rho-\omega-\sigma$ model because of the 
possibility of scalar-vector mixing the mixed polarization
tensor has the following form in ANM,
\begin{equation}
\Pi=\pmatrix{\Pi_{\rho \rho} & \Pi_{\rho \omega} & \Pi_{\rho \sigma} \cr 
\Pi_{\omega \rho} & \Pi_{\omega \omega} & \Pi_{\omega \sigma} \cr 
\Pi_{\sigma \rho} & \Pi_{\sigma \omega} & \Pi_{\sigma \sigma}}.  
\end{equation}
The propagator in the presence of mixining is the solution of
\begin{equation}
D^{-1} = D_0^{-1}-\Pi,
\end{equation}
where $D_0={\mathrm{diag}}(d_\rho, d_\omega, d_\sigma)$ with
$d_i=1/(Q^2-m_i^2+i\epsilon), i=\rho, \omega, \sigma$.

The mixed propagator can be obtained by matrix inversion:
\begin{equation}
D=
\left(
\begin{array}{lll}
 \frac{s_{\sigma } s_{\omega }-\Pi _{\sigma \omega } \Pi _{\omega \sigma }}{\Delta } & \frac{\Pi _{\rho \sigma } \Pi _{\sigma
   \omega }-s_{\sigma } \Pi _{\rho \omega }}{\Delta } & \frac{\Pi _{\rho \omega } \Pi _{\omega \sigma }-s_{\omega } \Pi _{\rho
   \sigma }}{\Delta } \\
 \frac{\Pi _{\sigma \rho } \Pi _{\omega \sigma }-s_{\sigma } \Pi _{\omega \rho }}{\Delta } & \frac{s_{\rho } s_{\sigma }-\Pi
   _{\rho \sigma } \Pi _{\sigma \rho }}{\Delta } & \frac{\Pi _{\rho \sigma } \Pi _{\omega \rho }-s_{\rho } \Pi _{\omega \sigma
   }}{\Delta } \\
 \frac{\Pi _{\sigma \omega } \Pi _{\omega \rho }-s_{\omega } \Pi _{\sigma \rho }}{\Delta } & \frac{\Pi _{\rho \omega } \Pi
   _{\sigma \rho }-s_{\rho } \Pi _{\sigma \omega }}{\Delta } & \frac{s_{\rho } s_{\omega }-\Pi _{\rho \omega } \Pi _{\omega
   \rho }}{\Delta }
\end{array}
\right),
\end{equation}
where $\Delta=s_{\rho } s_{\sigma } s_{\omega }-\Pi _{\rho \sigma } \Pi _{\sigma \rho } s_{\omega }-s_{\sigma } \Pi _{\rho \omega } \Pi
   _{\omega \rho }+\Pi _{\rho \sigma } \Pi _{\sigma \omega } \Pi _{\omega \rho }+\Pi _{\rho \omega } \Pi _{\sigma \rho } \Pi
   _{\omega \sigma }-s_{\rho } \Pi _{\sigma \omega } \Pi _{\omega \sigma }$.

The polarization functions in $D$ correspond to the mixing of the
pure isospin eigenstates, $\rho_I, \omega_I$ and $\sigma_I$. This matrix
can be expressed in terms of the propagators of the physical
$\rho, \omega$ and $\sigma$ fields which is an admixture of
the pure states. This is achieved by appropriate field rotation involving
various mixing angles in isospin space :
\begin{equation}
\pmatrix{\rho \cr \omega \cr \sigma} = {\tilde{S}}\,\pmatrix{\rho_I \cr \omega_I 
\cr \sigma_I},
\end{equation}
where $\tilde{S}$ is $3\times 3$ mixing matrix involving density
dependent mixing angles.
Let us first compare the polarization tensors $\Pi_{\omega \sigma}$ and 
$\Pi_{\rho \sigma}$ 
which have the same three momentum dependence. Now since $\Pi_{\omega \sigma}$
is a function of sum of the proton and neutron densities and $\Pi_{\rho \sigma}$
is that of the difference, thus $\Pi_{\omega \sigma} >> \Pi_{\rho \sigma}$
for small values of the asymmetry parameter $\alpha$. Taking 
$\Pi_{\rho \sigma} \approx 0$ the propagator matrix takes the form:

\begin{equation}
D=
\left(
\begin{array}{lll}
 \frac{s_{\sigma } s_{\omega }-\Pi _{\sigma \omega } \Pi _{\omega \sigma }}{\Delta_1 } & -\frac{s_{\sigma } \Pi _{\rho \omega
   }}{\Delta_1 } & \frac{\Pi _{\rho \omega } \Pi _{\omega \sigma }}{\Delta_1 } \\
 -\frac{s_{\sigma } \Pi _{\omega \rho }}{\Delta_1 } & \frac{s_{\rho } s_{\sigma }}{\Delta } & -\frac{s_{\rho } \Pi _{\omega
   \sigma }}{\Delta_1 } \\
 \frac{\Pi _{\sigma \omega } \Pi _{\omega \rho }}{\Delta_1 } & -\frac{s_{\rho } \Pi _{\sigma \omega }}{\Delta_1 } & \frac{s_{\rho }
   s_{\omega }-\Pi _{\rho \omega } \Pi _{\omega \rho }}{\Delta_1 }
\end{array}
\right),
\end{equation}
where
$\Delta_1=s_{\rho } s_{\sigma } s_{\omega }-s_{\sigma } \Pi _{\rho \omega } \Pi _{\omega \rho }-s_{\rho } \Pi _{\sigma \omega } \Pi
   _{\omega \sigma }$

Now it is also known that the 
scalar-vector mixing vanishes in the limit of vanishing 3-momentum
and is very small for low momentum. However, $\Pi_{\rho \omega}$ remains
non-zero even for $\vec q =0$. It is also to be noted that medium
dependent mixing effects are dominant only in the low momentum region which
allows us to put $\Pi_{\omega \sigma} \approx 0$ in this limit~\cite{wolf}.
Under these assumptions, $D$ reduces to block diagonal form
in which $\sigma$ decouples from $\rho$ and $\omega$ so that we have,  

\begin{equation}
D=\pmatrix{\frac{s_{\omega}}{s_{\omega}s_{\rho}-\Pi_{\rho \omega}^2} & 
\frac{\Pi_{\rho \omega}}{s_{\rho}s_{\omega}-\Pi_{\rho \omega}^2} & 0 \cr
\frac{\Pi_{\rho \omega}}{s_{\rho}s_{\omega}-\Pi_{\rho \omega}^2} & 
\frac{s_{\rho}}{s_{\rho}s_{\omega}-\Pi_{\rho \omega}^2} & 0 \cr
0 & 0 & 1/s_{\sigma}}.
\end{equation}

Consequently, $\tilde{S}$ takes the form

\begin{equation}
{\tilde{S}} = \pmatrix{\cos{\epsilon} & \sin{\epsilon} & 0 \cr
-\sin{\epsilon} & \cos{\epsilon} & 0 \cr
0 & 0 & 1}.
\end{equation}

We henceforth work in the $2\times 2$ subspace of $\rho$ and $\omega$,
i. e. we use :
\begin{equation}
S = \pmatrix{\cos{\epsilon} & \sin{\epsilon} \cr 
-\sin{\epsilon} & \cos{\epsilon}}.
\end{equation}
The mixing angle can be deduced by diagonalising the vector meson 
propagator~\cite{connell97}:
\begin{equation}
S^{-1}
\pmatrix{\frac{s_{\omega}}{s_{\omega}s_{\rho}-\Pi_{\rho \omega}^2} & \frac{\Pi_{\rho \omega}}{s_{\rho}s_{\omega}-\Pi_{\rho \omega}^2} \cr
\frac{\Pi_{\rho \omega}}{s_{\rho}s_{\omega}-\Pi_{\rho \omega}^2} & 
\frac{s_{\rho}}{s_{\rho}s_{\omega}-\Pi_{\rho \omega}^2}}
S
=
\pmatrix{\frac{s_{\omega}}{s_{\rho}s_{\omega}-\Pi_{\rho \omega}^2} & 0 \cr
0 & \frac{s_{\rho}}{s_{\rho}s_{\omega}-\Pi_{\rho \omega}^2}}, 
\end{equation}
from which we obtain
\begin{equation}
\tan{2\epsilon} = \frac{2\Pi_{\rho \omega}}{s_{\rho}-s_{\omega}}.
\end{equation}
In the small mixing angle limit the above equation reduces to
\begin{equation}
\epsilon = \frac{\Pi_{\rho \omega}}{s_{\rho}-s_{\omega}}.
\end{equation}
For the derivation of pion form factor for arbitrary mixing angle we need to
evaluate the matrix element for the process $\gamma \rightarrow \pi \pi$,
which can be written as 
\begin{eqnarray}
{\cal{M}}_{\gamma \rightarrow \pi \pi}^{\mu} 
&=&
\pmatrix{{\cal{M}}_{\rho_I \rightarrow \pi \pi}^{\mu} & 0}SS^{-1}
\pmatrix{\frac{s_{\omega}}{s_{\rho}s_{\omega}-\Pi_{\rho \omega}^2}
 & \frac{\Pi_{\rho \omega}}{s_{\rho}s_{\omega}-\Pi_{\rho \omega}^2} \cr
\frac{\Pi_{\rho \omega}}{s_{\rho}s_{\omega}-\Pi{\rho \omega}^2} &
 \frac{s_{\rho}}{s_{\rho}s_{\omega}-\Pi_{\rho \omega}^2}}\nonumber\\
&&\times\, SS^{-1}
\pmatrix{{\cal{M}}_{\gamma \rightarrow \rho_I} \cr
{\cal{M}}_{\gamma \rightarrow \omega_I}} 
\end{eqnarray}
where it is assumed that $\Gamma_{\omega_I \rightarrow \pi \pi} = 0$.
The physical amplitudes can be obtained from the above equation which are
as follows :
\begin{eqnarray}
{\cal{M}}_{\rho \rightarrow \pi \pi}^{\mu}&=&
\cos{\epsilon}{\cal{M}}_{\rho_I \rightarrow \pi \pi}^{\mu},\nonumber\\
{\cal{M}}_{\omega \rightarrow \pi \pi}^{\mu}&=&
\sin{\epsilon}{\cal{M}}_{\rho_I \rightarrow \pi \pi}^{\mu},\nonumber\\
{\cal{M}}_{\gamma \rightarrow \rho}&=&
\cos{\epsilon} {\cal{M}}_{\gamma \rightarrow \rho_I} -
\sin{\epsilon}{\cal{M}}_{\gamma \rightarrow \omega},\nonumber\\
{\cal{M}}_{\gamma \rightarrow \omega}&=&
\sin{\epsilon} {\cal{M}}_{\gamma \rightarrow \rho_I} +
\cos{\epsilon}{\cal{M}}_{\gamma \rightarrow \omega}
\end{eqnarray}

\begin{figure}[htb]
\begin{center}
\epsfig{figure=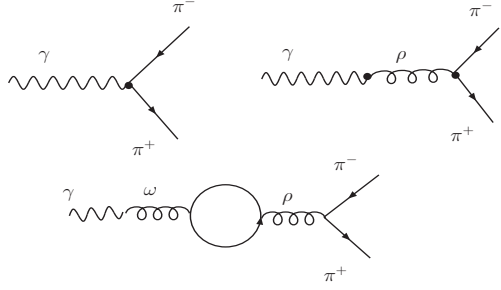, height=4.0 cm, width=7cm}
\caption{Contribution to $\rho$-$\omega$ mixing to pion form factor.}
\label{app_fig1}
\end{center}
\end{figure}

Thus the amplitude for $\gamma \rightarrow \pi \pi$ can be written
as
\begin{eqnarray}
{\cal{M}}_{\gamma \rightarrow \pi \pi}^{\mu}&=&
{\cal{M}}_{\rho \rightarrow \pi \pi}^{\mu}
\frac{s_{\omega}}{s_{\rho}s_{\omega}-\Pi_{\rho \omega}^2}
{\cal{M}}_{\gamma \rightarrow \rho}\nonumber\\
&&+
{\cal{M}}_{\rho \rightarrow \pi \pi}^{\mu}
\frac{s_{\rho}}{s_{\rho}s_{\omega}-\Pi_{\rho \omega}^2}
\nonumber\\
&&\times{\tan\epsilon}
{\cal{M}}_{\gamma \rightarrow \omega}
\end{eqnarray}
With these it is very easy to read out pion electromagnetic form factor from
Fig.(\ref{app_fig1}),
\begin{eqnarray}
F_{\pi}(Q^2)& =& 1-g_{\rho \pi \pi}\frac{s_\omega}{s_{\rho}s_{\omega}-\Pi_{\rho \omega}^2}\frac{Q^2}{g_{\rho}}
-g_{\rho \pi \pi} \tan{\epsilon}\nonumber\\
&&\times \frac{s_{\rho}}{s_{\rho}s_{\omega}-\Pi_{\rho \omega}^2}\frac{Q^2}{g_{\omega}}
\end{eqnarray}
which in the small angle limit reduces to Eq.(20).

%


\section*{Appendix - II}

In this appendix the mathematical expressions for polarization
functions are presented.
\bea
\Pi_{\mu\nu}^{vv}&=&\frac{g_ig_j}{\pi^3} 
\left [ A_{\mu\nu}-C_{\mu\nu} \right]\nn\\
\Pi_{\mu\nu}^{vt+tv}&=&-\frac{g_i g_j}{\pi^3} 
\frac{(\kappa_i + \kappa_j) M^\ast}{4M}B_{\mu\nu}\nn\\
\Pi_{\mu\nu}^{tt}&=&-\frac{g_i g_j}{\pi^3} \frac{\kappa_i\kappa_j}{4M^2} 
\left [  Q^2 A_{\mu\nu} + M^{\ast \ 2} B_{\mu\nu} \right ]
\eea

\bea
A_{\mu\nu}&=&Q^2\int \frac{d^3k}{E_k^\ast} \frac{{\cal{K}}_{\mu\nu}}
{Q^4-4(K\cdot Q)^2}\nn\\
&=&Q^2\,\int\,\frac{d^3k}{E_k^\ast}\,\frac{K_{\mu} K_{\nu}}{Q^4-4(K\cdot Q)^2}
+\frac{Q_{\mu} Q_{\nu}}{4Q^2}({\tilde{B}}-\alpha(k_f))\nn\\
&&-\frac{1}{4}\int\frac{d^3k}{E_k^\ast} (Q_{\mu}K_{\nu}+Q_{\nu}K_{\mu})\,
\left[\frac{1}{Q^2-2K\cdot Q}\right.\nn\\
&&-\left.\frac{1}{Q^2+2K\cdot Q}\right]\nn\\ 
B_{\mu\nu}&=&Q^4\int \frac{d^3k}{E_k^\ast} 
\frac{Q_{\mu\nu}} {Q^4-4(K\cdot Q)^2}\nn\\
C_{\mu\nu}&=&\int \frac{d^3k}{E_k^\ast} \frac{ (K\cdot Q)^2 Q_{\mu\nu}}
{Q^4-4(K\cdot Q)^2}
\eea

\bea
%
%
A_{T}&=& Q^2\int \frac{d^3k}{E_k^\ast}\frac{k_T^2}{Q^4-4(K\cdot Q)^2}\nn\\ 
&=&\frac{\pi}{4q}
\int \frac{k^3dk}{\Ekst} \cI_2\nn\\
&&-\frac{\pi}{16q^3}\int \frac{kdk}{\Ekst} 
\left [\frac{}{}(Q^4+4E_k^{\ast\ 2}\om^2)\cI_2-4Q^2\Ekst\om\cI_1\right.\nn\\
&&-\left. 8Q^2kq\right ]
\eea

\bea
A_{L}&=&Q^2 \int \frac{d^3k}{E_k^\ast} \frac{k_z^2-E_k^{ \ast 2}}{Q^4
-4(K\cdot Q)^2} 
- \frac{1}{4}( {\tilde{B}}-\alpha(k_f) )\nn\\ 
&&+\frac{1}{2} \int \frac{d^3k}{E_k^\ast}  (\omega E_k^\ast - q_z k_z)\nn\\
&&\times\left 
[ \frac{1}{Q^2 - 2 K\cdot Q} -   \frac{1}{Q^2 + 2 K\cdot Q}    \right]\nn\\
&=&
\frac{\pi}{8q^3}\int \frac{kdk}{\Ekst} 
\left [(Q^4+4E_k^{\ast\ 2}\om^2)\cI_2-4Q^2\Ekst\om\cI_1\right.\nn\\
&&\left.-8Q^2kq\right ]-\frac{\pi}{2q}\int kdk E_K^\ast \cI_2\nn\\
&&+\frac{\pi}{4q}\int \frac{kdk}{E_k^\ast} [Q^2\cI_2-8kq]-\frac{1}{4}
 [ {\tilde{B}}-\alpha(k_f) ]
\eea

\bea
B_{T}&=& B_{L} =B , C_{T}= C_{L} =C
\eea

\bea
B&=& 
Q^4\int \frac{d^3k}{E_k^\ast}\frac{1}{Q^4-4(K\cdot Q)^2}\nn\\
&=&\frac{\pi Q^2}{2q}\int \frac{kdk}{E_k^\ast}\cI_2\nn\\ 
C&=& \int \frac{d^3k}{E_k^\ast}\frac{(K\cdot Q)^2}
{Q^4-4(K\cdot Q)^2}\nn\\
&=&\frac{\pi}{8q}\int \frac{kdk}{E_k^\ast} [Q^2\cI_2-8kq]\nn\\
\alpha(k_f) &=&\int_0^{k_f}\,\frac{d^3k}{E_k^\ast}\nn\\ 
&=&2\pi \left[k_f\epsilon_f - M^{\ast 2}
\ln\frac{k_f+\epsilon_f}{M^{\ast }}\right]\nn\\
{\tilde{B}}&=&\frac{\pi Q^2}{2q}\int \frac{kdk}{E_k^\ast} \cI_1
\eea

and 
\bea
\cI_1&=&\ln \frac{Q^4-4(E_k^\ast \omega - kq)^2}{Q^4-4(E_k^\ast \omega + kq)^2}
\nn\\
\cI_2&=&
\ln\frac{(Q^2+2k q)^2-4 E_k^{\ast 2}\omega^2 }{(Q^2-2k q)^2
-4 E_k^{\ast 2}\omega^2 } 
\eea

\end{document}